\documentclass[]{cupconf}
\usepackage{graphicx}

\title[Oxygen-rich droplets and the enrichment of the ISM]
      {Oxygen-rich droplets and the enrichment of the ISM}
      
\author[Stasi\'nska et al.]{\\ Gra\.zyna Stasi\'nska$^1$, Guillermo Tenorio-Tagle$^2$, M\'onica Rodr\'\i guez$^2$, William J. Henney$^3$}

\affiliation{$^1$ LUTH, Observatoire de Paris-Meudon, 5 Place Jules Jansen, 92195 Meudon,   France \\ $^2$ Instituto Nacional de Astrof\'\i sica \'Optica y Electr\'onica, AP 51, 72000, Puebla, Mexico \\ $^3$Centro de Radioastronom\'{\i}a y Astrof\'{\i}sica, Universidad Nacional Aut\'onoma
de M\'exico, Campus Morelia, Apartado Postal 3-72, 58090 Morelia, Mexico }   

\begin{document}
\maketitle

\begin{abstract}
We argue that the discrepancies observed in HII regions  between  abundances derived from optical recombination lines (ORLs) and collisionally excited lines (CELs) might well be the signature of a scenario of the enrichment of the interstellar medium (ISM) proposed by Tenorio-Tagle (1996). In this scenario, the fresh oxygen released during massive supernova explosions is confined within the hot superbubbles as long as supernovae continue to explode. Only after the last massive supernova explosion, the metal-rich gas starts cooling down and falls on the galaxy within metal-rich droplets.  Full mixing of these metal-rich droplets and the ISM occurs during photoionization by the next generations of massive stars. During this process, the metal-rich droplets give rise to strong recombination lines of the metals, leading to the observed ORL-CEL discrepancy. (The full version of this work is submitted to Astronomy and Astrophysics.)
\end{abstract}

\section{Introduction}

There is no doubt that galaxies suffer chemical enrichment during their lives (see e.g. Cid Fernandes et al. 2006 for a recent systematic approach using a large data base of galaxies from the Sloan Digital Survey Data Release 5 -- Adelman-Mac Carthy  J.K. et al., 2007). The main source of oxygen production has since long been identified as due to supernovae from massive stars (type II supernovae). 
Yet, the exact process by which chemical enrichment proceeds is poorly known (see a review by Scalo \& Elmegreen 2004). 

Ten years ago, Tenorio-Tagle (1996, hereafter T-T96) proposed a scenario in which the metal-enhanced ejecta from supernovae follow a long excursion in galactic haloes before falling down on the galaxies in the form of oxygen-rich droplets.

\begin{figure} 
      \centering
\includegraphics[width=4in,height=2in]{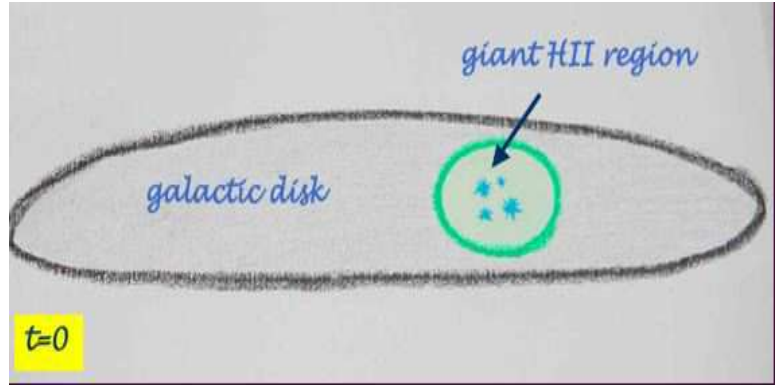}
\caption{Sketch of the T-T96 scenario: At time t=0, a burst of star formation occurs and a giant HII region forms.}\label{fig}
\end{figure}

\begin{figure}  
      \centering
\includegraphics[width=3in,height=3in]{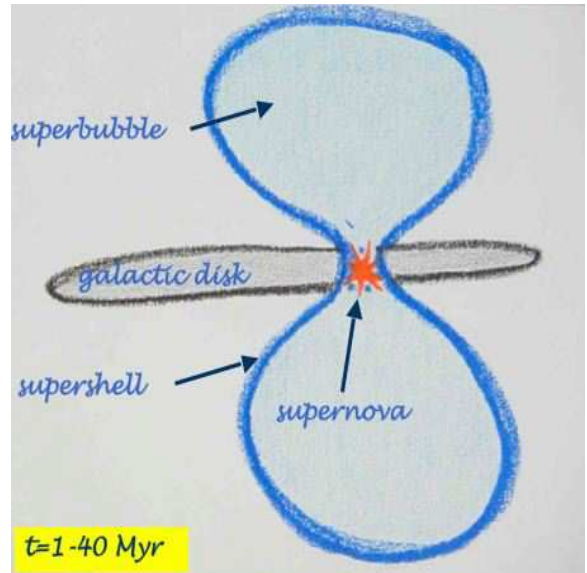}
\caption{Sketch of the T-T96 scenario: During the next $\sim$ 40~Myr, supernovae explode, creating a hot superbubble confined within a large expanding  supershell that bursts into the galactic halo. The  superbubble contains the matter from the oxygen-rich supernova ejecta mixed with the matter from the stellar winds and with the matter thermally evaporated from the surrounding supershell.}\label{fig}
\end{figure}

In the present work (the full version of which has been submitted to Astronomy \& Astrophysics, Stasi\'nska et al. 2007), we suggest that the discrepancy between the oxygen abundances derived from optical recombination lines (ORLs) and from collisionally excited lines (CELs) in HII regions (see e.g. Garc\'\i a-Rojas et al. 2006 and references therein) might well be the signature of those oxygen-rich droplets. In fact, Tsamis et al. (2003) and P\'equignot \& Tsamis (2005) already suggested that the ORL/CEL discrepancy in HII regions is the result of inhomegeneities in the chemical composition in these objects.
Our aim is to explicit the link between the ORL/CEL discrepancy and the T-T96 scenario, and to check whether what is known of the oxygen yields allows one to explain the ORL/CEL discrepancy in a quantitative way.

\section{The Tenorio-Tagle (1996) scenario}

Figures 1--5 present the T-T96 scenario  in cartoon format.

\begin{figure}  
      \centering
\includegraphics[width=3in,height=3in]{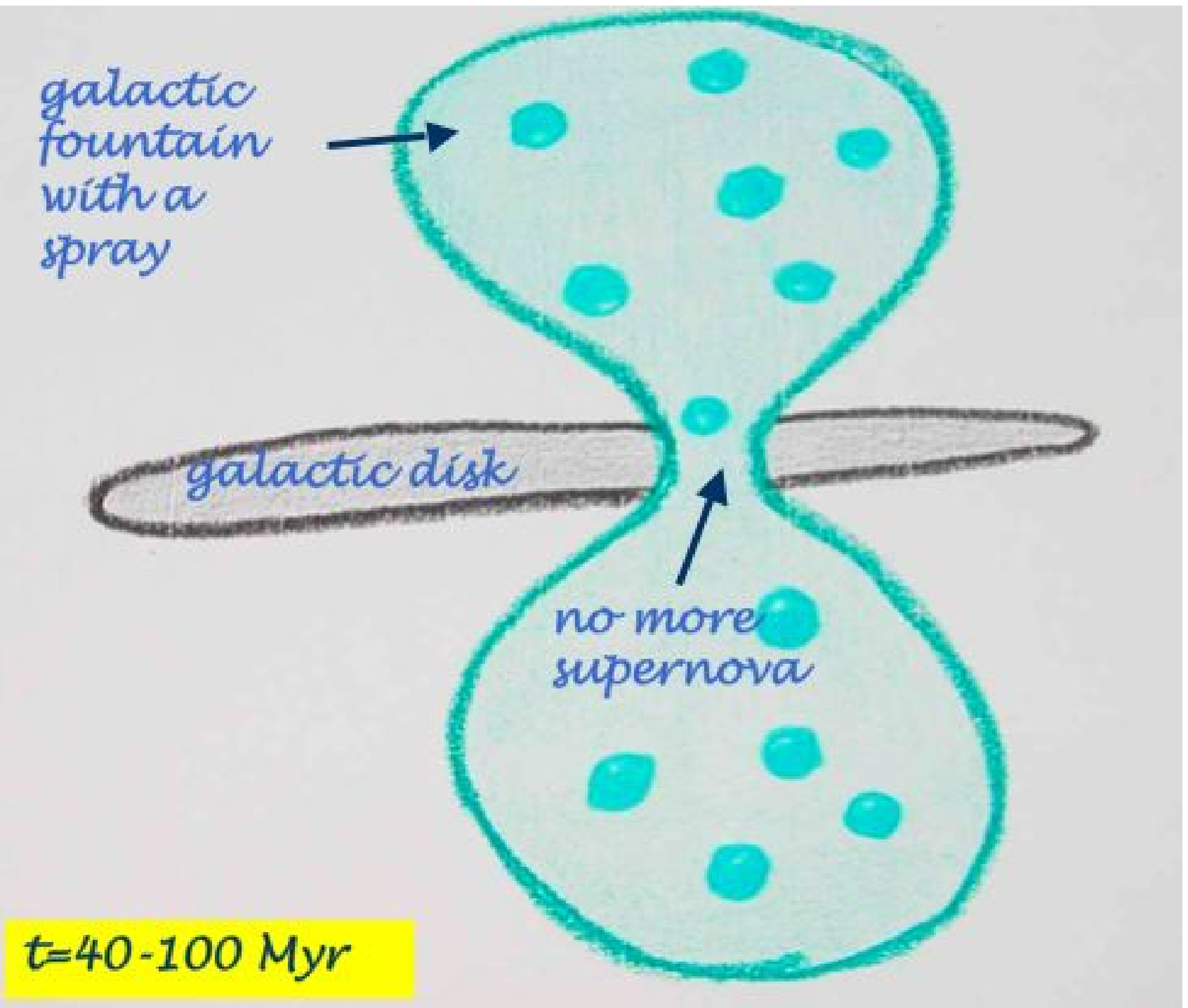}
\caption{Sketch of the T-T96 scenario: after the last supernova has exploded, the gas in the superbubble begins to cool down. Loci of higher densities cool down quicker. Due to a sequence of fast repressurizing shocks, this leads to the formation of metal-rich cloudlets. The cooling timescale is of the order of 100~Myr.}\label{fig}
\end{figure}

\begin{figure}  
      \centering
\includegraphics[width=3in,height=2.6in]{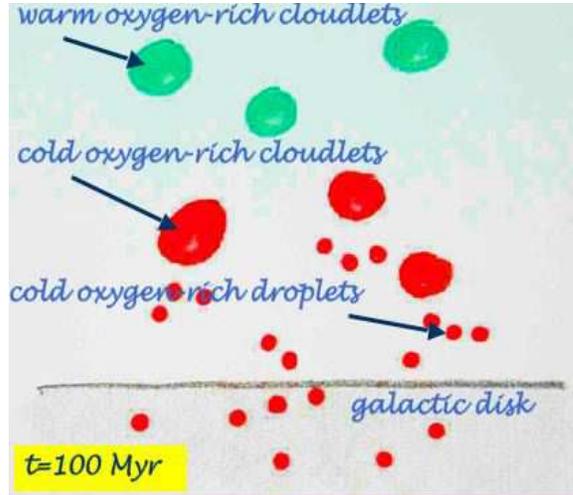}
\caption{Sketch of the T-T96 scenario: The now cold metal-rich cloudlets fall unto the galactic disk. They are further fragmented into metal-rich droplets by Raighleigh-Taylor instabilities. This metal-rich rain affects a region whose extension is of the order of kiloparsecs, i.e. much larger than the size of the initial HII region}\label{fig}
\end{figure}

\begin{figure} 
      \centering
\includegraphics[width=3.5in,height=3.5in]{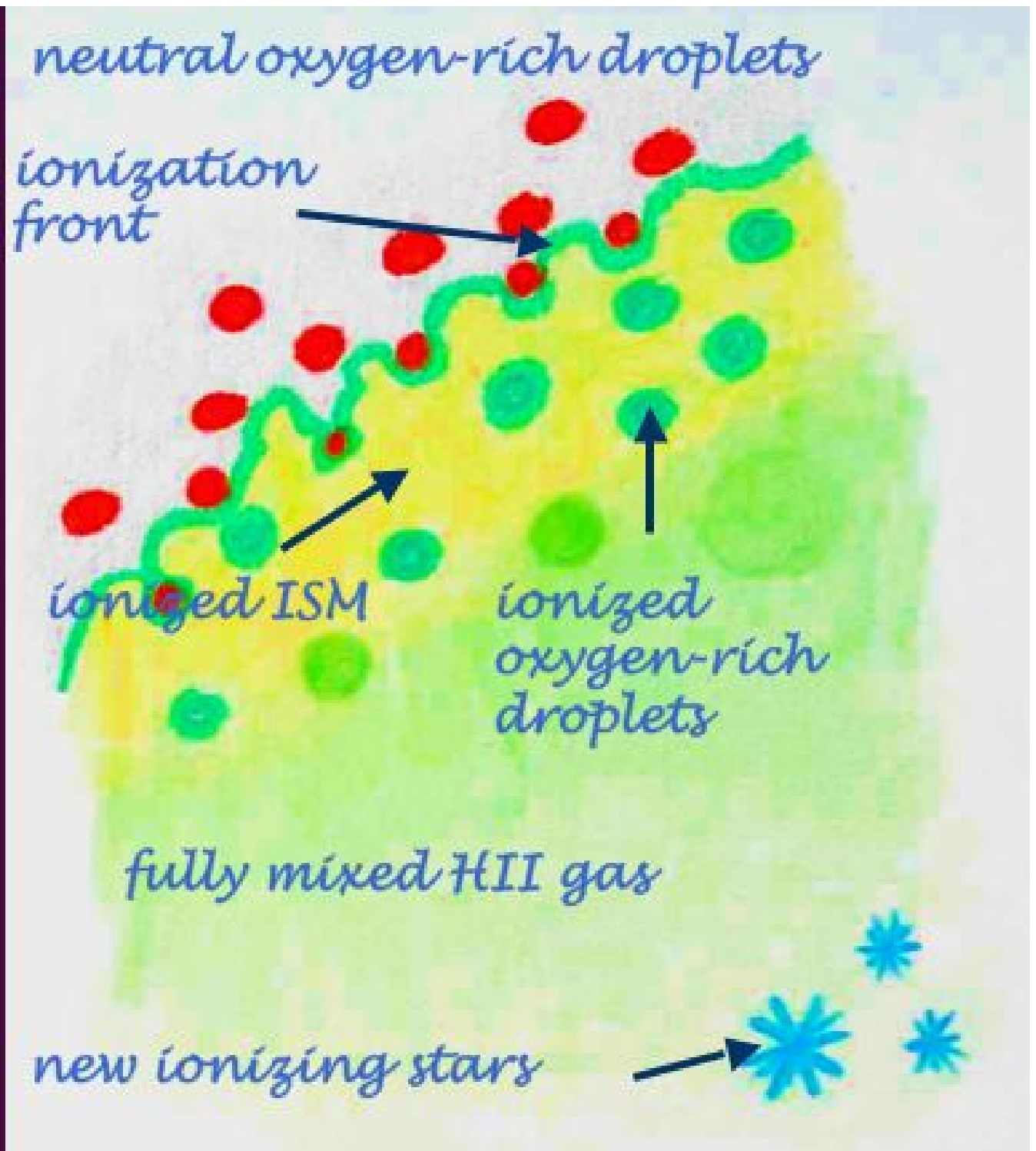}
\caption{Sketch of the T-T96 scenario: When a next generation of massive stars form, they photoionize the surrounding interstellar medium, including the metal-rich droplets. It is only after the droplets have been photoionized that their matter is intimately mixed with the matter from the ISM, and that proper chemical enrichment has occured. The whole process since the explosion of the supernovae that provided fresh oxygen has taken at least 100~Myr. }\label{fig}
\end{figure}

\section{The ORL-CEL discrepancy in the context of the T-T96 scenario} 

The details of the physical arguments concerning the amount of oxygen available in the droplets, the mixing processes, as well as the simulation of the ORL-CEL discrepancy with a multizone photoionization model are described in Stasi\'nska et al. (2007). Here, we simply give the most important conclusions.

Photoionization of the oxygen-rich  droplets predicted by the T-T96 scenario can reproduce the
observed abundance discrepancy factors (ADFs, i.e. the ratios of abundances obtained from ORLs and from CELs) 
derived for Galactic and extragalactic HII
regions.
The recombination lines arising from the highly 
metallic droplets thus show mixing  at work.

We find that, if our scenario holds,  the 
recombination lines strongly overestimate the 
metallicities of the fully mixed HII regions. 
The collisionally excited lines may also 
overestimate them, although in much smaller 
proportion.   In absence of any 
recipe to correct for these biases, we recommend 
to discard objects showing large ADFs to probe 
the chemical evolution of galaxies.

To proceed further with this question of 
inhomogeneities, one needs as many 
observational
constraints as possible.

On the theoretical side, one needs more 
robust estimates of the integrated stellar yields 
as well as a better knowledge of the impact 
of massive stars on the ISM and of the role of turbulence.
All these issues are relevant to our understanding of the metal enrichment of the Universe.

\end{document}